\documentclass[twocolumn,showpacs,preprintnumbers,amsmath,amssymb]{revtex4}

\usepackage{graphicx}
\usepackage{dcolumn}
\usepackage{bm}

\begin{document}

\preprint{APS/123-QED}

\title{6$^+$ isomers in neutron-rich Sn-isotopes beyond N$=82$ and effective interaction}
\author{Bhoomika Maheshwari}
\email{bhoomika.physics@gmail.com}
\author{Ashok Kumar Jain}%
\author{P. C. Srivastava}
\affiliation{Department of Physics, Indian Institute of Technology, Roorkee 247667, India.}
\date{\today}

\begin{abstract}
Recent observation of the 6$^+$ seniority isomers and measurements of the B${(E2)}$ values in the $^{134-138}$Sn isotopes lying close to the neutron drip line have raised some questions about the validity of the currently used effective interactions in the neutron-rich region. Simpson $\it{et}$ $\it{al.}$ [Phys. Rev. Lett. $\textbf{113}$, 132502 (2014)] had to modify the diagonal and non-diagonal ${\nu}f_{7/2}^2$ two body matrix elements of the ``${Vlk}$'' interaction by ${\sim150}$ keV in their shell model calculations in order to explain the data of $^{136}$Sn. In contrast, we are able to explain the observed energy levels and the B${(E2)}$ values after marginal reduction of the same set of matrix elements by ${25}$ keV in the ``RCDB'' (Renormalized CD-Bonn) interaction. The observed mismatch in reproducing the data of $^{136}$Sn is due to the seniority mixing. Further, we do not find it necessary to consider the core excitations, and the ``RCDB'' interaction seems better suited to explain the data beyond N$=82$ magic number.
\end{abstract}

\maketitle

\section{\label{sec:level1}Introduction}

The seniority isomers~\cite{talmi} in the semi-magic nuclei present a crucial testing ground for the large scale shell model calculations as well as the effective interactions. The Z${=50}$ Sn-isotopes, known from $^{100}$Sn to $^{134}$Sn, have been the focus of a large number of recent studies. In this paper, we focus upon the neutron-rich Sn-isotopes beyond the N${=82}$ closed shell, where new data have recently become available in $^{136,138}$Sn~\cite{simpson}.  More specifically, the 6$^+$ isomers in $^{134-138}$Sn-isotopes present an interesting set of data to test the validity of the effective interactions in neutron-rich systems close to the drip line. Some of the earlier works in the Sn-isotopes beyond A${=132}$ are due to Hoff $\it{et}$ $\it{al.}$~\cite{hoff} in $^{133}$Sn, Zhang $\it{et}$ $\it{al.}$~\cite{zhang}, Korgul $\it{et}$ $\it{al.}$~\cite{korgul} and Beene $\it{et}$ $\it{al.}$~\cite{beene} in $^{134}$Sn. Theoretical works on the effective interactions and the shell model calculations beyond N$=82$ have also been reported by Kartamyshev $\it{et}$ $\it{al.}$~\cite{kartamyshev}, Sarkar and Sarkar~\cite{sarkar}, and Covello $\it{et}$ $\it{al.}$~\cite{covello}.

In a recent experiment, Simpson $\it{et}$ $\it{al.}$~\cite{simpson} have populated the $^{136,138}$Sn-isotopes by using the RIBF facility at RIKEN. They have reported measurements of the low-lying energy levels as well as the life-times of the 6$^+$ isomers. They have combined the earlier studies on $^{134}$Sn, carried out by Zhang $\it{et}$ $\it{al.}$~\cite{zhang}, Korgul $\it{et}$ $\it{al.}$~\cite{korgul} and Covello $\it{et}$ $\it{al.}$~\cite{covello}, with their studies, and discuss the properties of the 6$^+$ isomers in the Sn-isotopes beyond the doubly magic $^{132}$Sn. They have also reported the results of large scale shell model calculations by using the ``OSLO'' code~\cite{engeland} along with the ``${Vlk}$'' interaction~\cite{bogner}. The calculations were able to reproduce the measured B$(E2)$ values reasonably well, except for $^{136}$Sn, where the calculated B$(E2; 6^+ \rightarrow 4_1^+)$ value between the yrast 6$^+$ and 4$^+$ levels differs from the experimental value by a factor of more than ${10}$. Their calculated B$(E2)$ value for $^{136}$Sn is very small and close to the value obtained from the pure $\it{v}$${=2}$ seniority scheme. They also found a mismatch between the calculated and the measured level energies of the yrast 2$^+$, 4$^+$ and 6$^+$ states for the $^{134-138}$Sn-isotopes. Further, they could not reproduce the measured B$(E2;6^+ \rightarrow4_1^+)$ value of $^{136}$Sn, even when the core excitations were included.

Simpson $\it{et}$ $\it{al.}$~\cite{simpson} were, however, able to explain the measured B$(E2)$ value of $^{136}$Sn after reducing the diagonal and non-diagonal ${\nu}f_{7/2}^2$ matrix elements by $\sim150$ keV. This has the effect of lowering the second 4$^+$ state (seniority $\it{v}$$=4$) by $250$ keV, bringing it much closer to the yrast 4$^+$ state (seniority $\it{v}$${=2}$). As a result, the two 4$^+$ states acquire almost $50\%$ mixing in seniority. This modified interaction also improves the level energies for all the three Sn-isotopes. The authors~\cite{simpson} conclude that further theoretical work is needed for constructing the effective interactions in the neutron-rich region. This has motivated us to test the new experimental data with other effective interactions in this region.

We have, therefore, carried out large scale shell model calculations by using the Nushell code of Brown and Rae~\cite{brown1} along with the ``RCDB'' interaction~\cite{brown2}, sometimes also referred as the ``CWG'' interaction. We find that we are able to explain the observed energy levels and the B$(E2)$ values quite well after reducing the ${\nu}f_{7/2}^2$ matrix elements by ${25}$ keV only in the ``RCDB'' interaction. Our results with the modified interaction may be helpful in further works in this direction. We present the details of the calculations in section II and the results in section III. The last section concludes the present work.

\section{\label{sec:level2}Calculations}

\begin{figure*}[!ht]
\includegraphics[width=14cm,height=11.5cm]{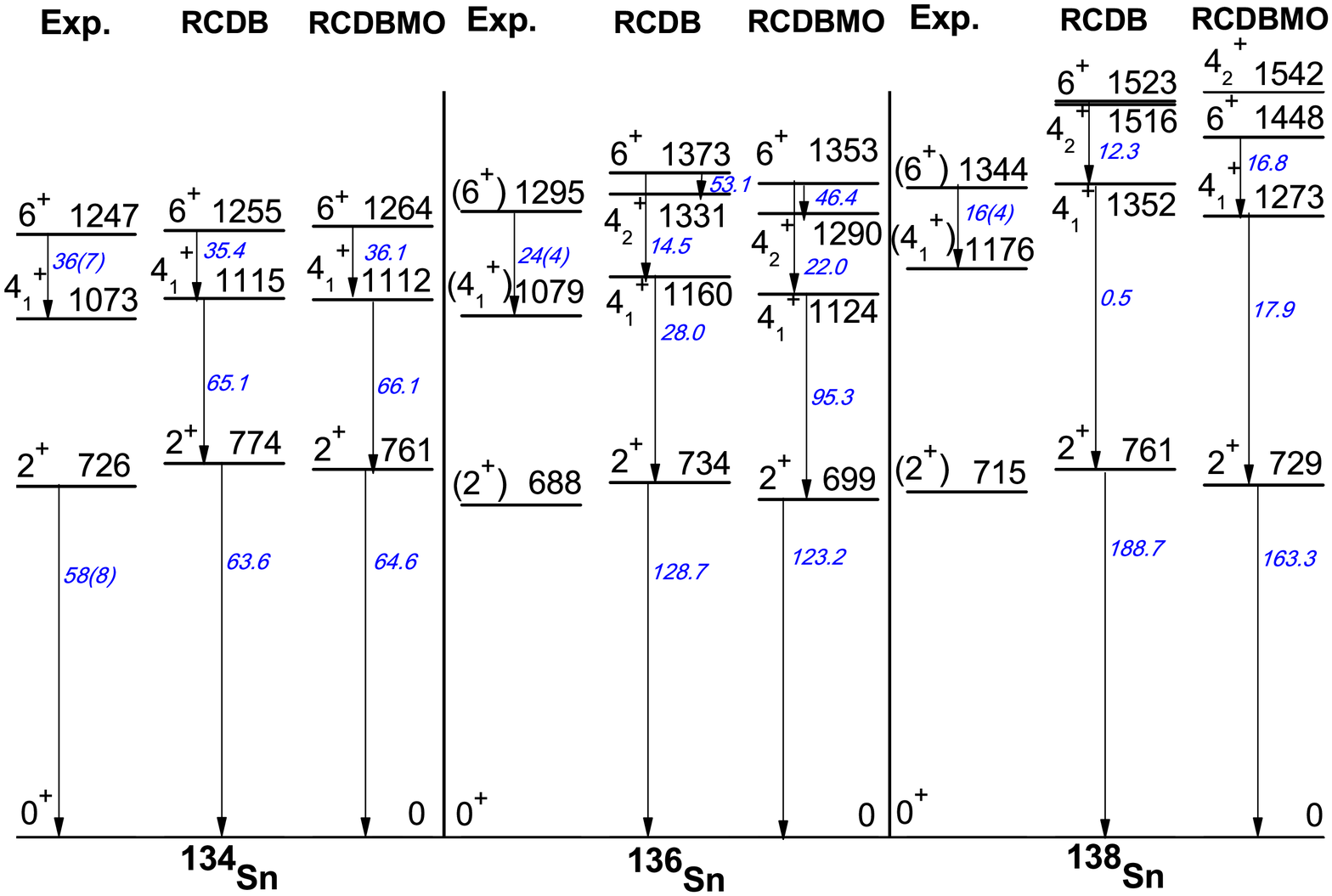}
\caption{\label{fig:level}(Color online) Energy level schemes for $^{134-138}$Sn. The experimental data shown on the left, are taken from~\cite{simpson,zhang,korgul,beene} and results of the shell model calculations with ``RCDB'' interaction and a modified ``RCDB'' (``RCDBMO'') interaction (see text for details) are shown on the right. The second $4_2^+$ state has been shown only for $^{136,138}$Sn and lies above the yrast $4_1^+$ state. All the energies are in keV. The B$(E2)$ values in units of $e^2fm^4$, shown in $\it{italics}$, have been rounded off.}
\end{figure*}

We have used the Nushell code~\cite{brown1} along with the ``RCDB'' interaction~\cite{brown2}, which has been modified for the latest neutron single particle energies~\cite{jones}, for calculating the spectroscopic properties of $^{134-138}$Sn. This interaction assumes the doubly-closed Z${=50}$, N${=82}$, $^{132}$Sn as an inert core and has been optimized for Z${=50-82}$ and N${=82-126}$ particle numbers. The harmonic oscillator potential was chosen with an oscillator parameter $\hbar \omega = 45A^{-1/3}-25A^{-2/3}$. The valence space consists of the six neutron $[2f_{7/2}, 3p_{3/2}, 1h_{9/2}, 3p_{1/2}, 2f_{5/2}, 1i_{13/2}]$ orbitals. The single particle energies for these neutron orbitals are ${-2.4550, -1.6010, -0.8940, -0.7990, -0.4500, 0.2500}$ MeV, respectively. These Sn-isotopes have ${2-6}$ valence neutrons so that only the neutron-neutron part of the interaction plays the major role. We have used the effective neutron charge to be $0.65$$\it{e}$ similar to the value used by Simpson $\it{et}$ $\it{al.}$~\cite{simpson}.

Our calculations are able to reproduce the experimental B$(E2; 6^+ \rightarrow 4_1^+)$ values for $^{134}$Sn and $^{138}$Sn reasonably well, but fail for $^{136}$Sn. Our calculated B$(E2)$ value for $^{136}$Sn differs from the experimental value by a factor of less than $2$. This, however, confirms the claim of ref.~\cite{simpson} that there is indeed a problem in explaining the results of $^{136}$Sn.

Following the prescription of Simpson $\it{et}$ $\it{al.}$~\cite{simpson}, we also decided to reduce the ${\nu}f_{7/2}^2$ diagonal and non-diagonal matrix elements. We find that it is possible to obtain very good results for both the level energies as well as the B$(E2)$ values of $^{134-138}$Sn after reducing the ${\nu}f_{7/2}^2$ matrix elements by a small amount of ${25}$ keV only. The calculated B$(E2)$ values now match extremely well with the experimental data. We denote these calculations with the modified interaction as ``RCDBMO''. The agreement of the calculated level schemes with the experimental levels also improves with ``RCDBMO'' as compared to the unmodified ``RCDB'' interaction.

\section{\label{sec:level3}Results and Discussion}

We have plotted in Fig.~\ref{fig:level}, the calculated and the experimental level  energies for the yrast 0$^+$, 2$^+$, 4$_1^+$ and 6$^+$ states in $^{134-138}$Sn-isotopes. The second $4_2^+$ state has also been shown only for $^{136,138}$Sn-isotopes. The figure also lists the measured B$(E2)$ values in units of $e^2 fm^4$ for all the transitions where experimental data exist. The experimental data for $^{134}$Sn have been taken from Zhang $\it{et}$ $\it{al.}$~\cite{zhang}, Korgul $\it{et}$ $\it{al.}$~\cite{korgul} and Beene $\it{et}$ $\it{al.}$~\cite{beene}, while those for $^{136,138}$Sn have been taken from Simpson $\it{et}$ $\it{al.}$~\cite{simpson}. We find that the calculated energies from the ``RCDB'' interaction, plotted in the middle for each isotope, generally lie higher than the measured ones.

We have also plotted in the last columns of Fig.~\ref{fig:level}, the calculated level energies obtained by using the modified interaction ``RCDBMO'', wherein the diagonal and non diagonal ${\nu}f_{7/2}^2$ matrix elements have been reduced by $25$ keV. We find that the modified interaction ``RCDBMO'' significantly improves the agreement between the calculated and the experimental level energies for $^{136,138}$Sn, while it leaves the levels of $^{134}$Sn almost unchanged. The second $4_2^+$ state in $^{136}$Sn comes down in energy by $41$ keV and lies below the yrast $6^+$ state. On the other hand, the second $4_2^+$ state in $^{138}$Sn goes up in energy and lies above the yrast $6^+$ state.

\begin{figure}[!ht]
\includegraphics[width=9cm,height=7cm]{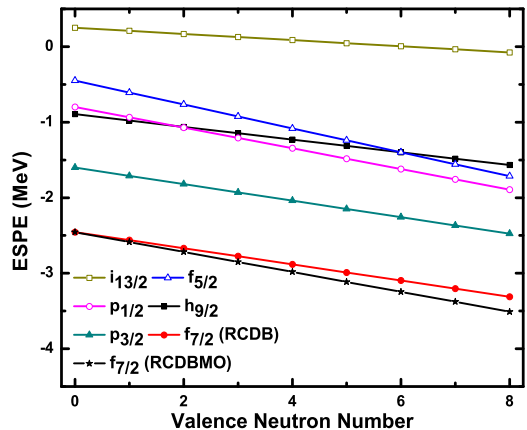}
\caption{\label{fig:espe}(Color online) Evolution of effective single particle energies of different orbitals with increasing neutron number for the ``RCDB'' interaction. The $f_{7/2}$ orbital is also plotted for the ``RCDBMO'' interaction. $^{132}Sn$ is taken as the core.}
\end{figure}

We have plotted in Fig.~\ref{fig:espe}, the effective single particle energies $(ESPE)$ of the valence neutron orbitals from valence neutron number $0$ to $8$ for the ``RCDB'' interaction. The $f_{7/2}$ orbital is also plotted for the ``RCDBMO'' interaction. The calculations have been done by using the relation ~\cite{otsuka},
\begin{equation}
\centering
ESPE = E_{j_{n}}+\sum_{j_n} {\bar{E}(j_n,j_n)}{\hat{n}}_{j_{n}}
\end{equation}
where $E_{j_{n}}$ and ${\hat{n}}_{j_{n}}$ denote the single-particle energies and number of neutrons occupying the orbital $j_{n}$. The term ${\bar{E}(j_n,j_n)}$ represents the monopole corrected interaction energy which has been averaged over the total angular momentum J, and is given by:
\begin{equation}
\centering
{\bar{E}(j_n,j_n)} = \cfrac{\sum_{J} (2J+1) <j_n j_n; J|V|j_n j_n; J>}{\sum_{J} (2J+1)}
\end{equation}
where $<j_n j_n; J|V|j_n j_n; J>$ stands for a two-body matrix element of the effective interaction. On using the modified interaction ``RCDBMO'', the $ESPE$ of the $f_{7/2}$ orbital get reduced by an amount which varies linearly with the particle number, while the $ESPE$ of rest of the orbitals remain the same. The small change in the energy of $f_{7/2}$ orbital plays a key role in deciding the calculated level energies of the yrast states in these nuclei. Even with this small change, our calculated energies come closer to the experimental ones.

\begin{figure}[!ht]
\includegraphics[width=9cm,height=7cm]{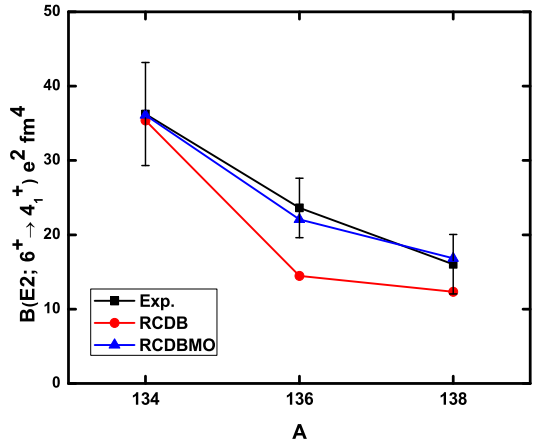}
\caption{\label{fig:be2}(Color online) Reduced transition rates for the yrast $6^+\rightarrow4_1^+$ transitions in $^{134-138}$Sn. Experimental data and error bars are taken from~\cite{simpson,zhang}. Calculated values for both ``RCDB'' and ``RCDBMO'' interactions are also shown.}
\end{figure}

We have also calculated the B$(E2)$ values for the transition between the yrast $(6^+ \rightarrow 4_1^+)$ states for $^{134-138}$Sn-isotopes and the same are plotted in Fig.~\ref{fig:be2}. The B$(E2)'$s for all the transitions between the yrast states have also been calculated and the rounded off values are shown in italics in Fig.~\ref{fig:level}. As already pointed out, our calculated B$(E2)$ values with the ``RCDB'' interaction reproduce the measured ones except for $^{136}$Sn isotope, where a deviation by a factor $<2$ is noticed. We have also calculated the B$(E2; 6^+\rightarrow4_2^+)$ value in $^{136}$Sn-isotope with both the ``RCDB'' and ``RCDBMO" interactions, and the rounded off values have been included in Fig.~\ref{fig:level}.

Semi-magic nuclei have been shown to be good candidates for the seniority isomerism because the $E2$ transition probability between the same seniority states become very small or vanish when the valence shell is close to half-filled. This is due to the fact that the matrix elements of even tensor operators between states with the same seniority vanish at the mid-shell~\cite{talmi,isacker}. Therefore, the B$(E2;6^+\rightarrow4_1^+)$ between the yrast $6^+$ and $4_1^+$ states, and B$(E2; 4_1^+ \rightarrow 2^+)$ between the yrast $4_1^+$ and $2^+$ states should diminish for $^{136}$Sn, a mid-shell nucleus, if these states have pure seniority $\it{v}$$=2$. Our calculated nonzero B$(E2)$ values for both the decays confirm that a mixing of $\it{v}$$=2$ and $\it{v}$$=4$ seniority is already present in $^{136}$Sn even with the unmodified ``RCDB'' interaction. In comparison, these states appear to have a pure $\it{v}$$=2$ seniority in the work of Simpson $\it{et}$ $\it{al.}$~\cite{simpson}. As a result, our calculated B$(E2)$ value in $^{136}$Sn is off by a factor $<2$, while Simpson $\it{et}$ $\it{al.}$~\cite{simpson} find it off by a factor $>10$.

\begin{figure}[!ht]
\includegraphics[width=9cm,height=8cm]{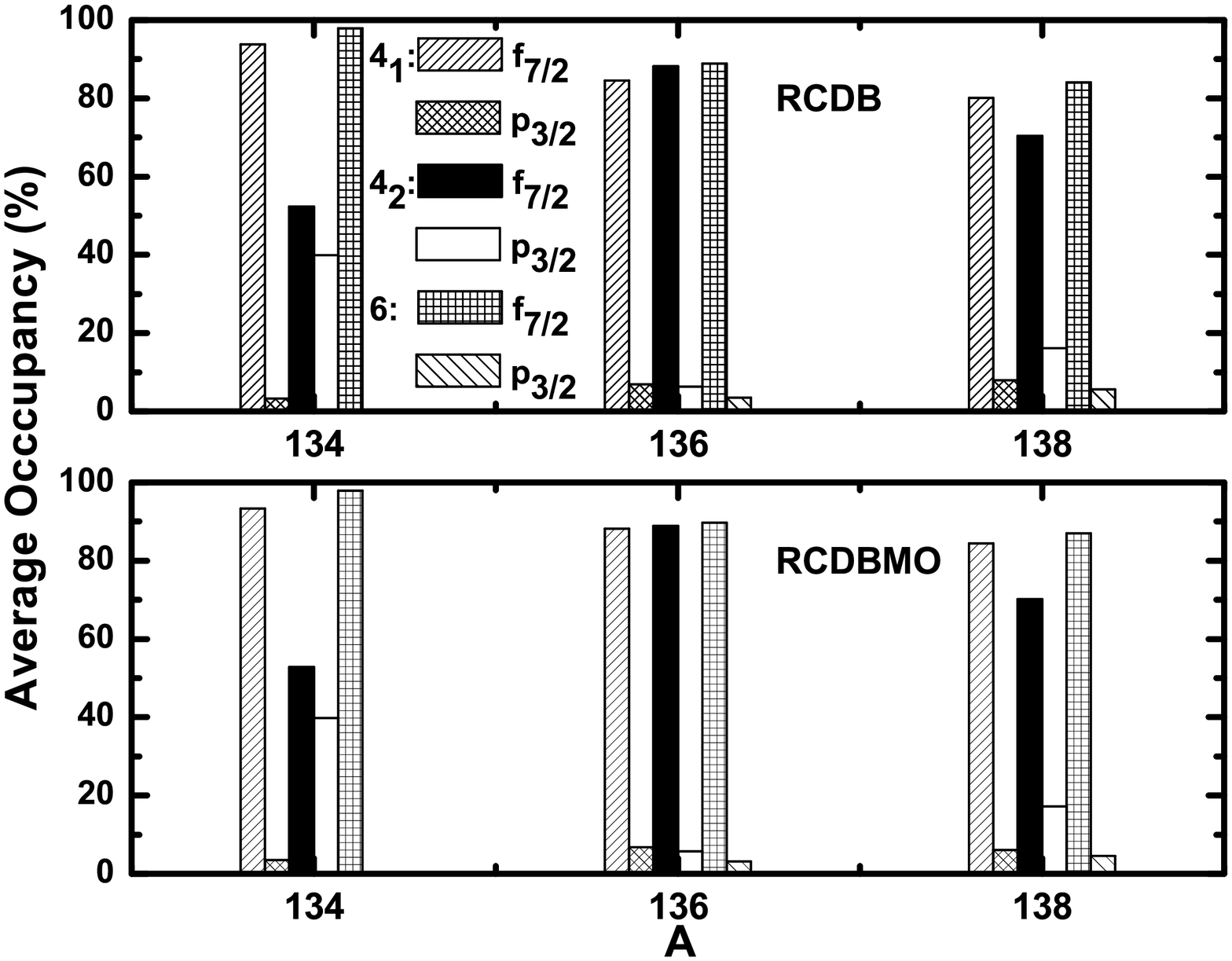}
\caption{\label{fig:occu} Average occupancy (in percentage) of the $f_{7/2}$ and $p_{3/2}$ orbitals for the ${4}_1^+$, $4_2^+$ and $6^+$ states in $^{134-138}$Sn-isotopes. }
\end{figure}

We find that the agreement for $^{136}$Sn in particular and all the three Sn-isotopes in general, significantly improves after reducing the ${\nu}f_{7/2}^2$ matrix elements by $25$ keV, a value much smaller than $\sim 150$ keV used by Simpson $\it{et}$ $\it{al.}$~\cite{simpson}. Our calculated B$(E2; 6^+\rightarrow4_1^+)$ value in $^{136}$Sn increases from $14.5$ $e^2fm^4$ to $22.0$ $e^2fm^4$, which comes quite close to the experimental value of $24(4)$ $e^2fm^4$. The calculated B$(E2)$ value between the yrast $4_1^+$ and $2^+$ states also increases from $28.0$ $e^2fm^4$ to $95.3$ $e^2fm^4$. This suggests that the seniority mixing for the yrast $4_1^+$ state has increased. The yrast $2^+$ and $6^+$ states remain dominantly seniority $\it{v}$$=2$ states as the second $2^+$ and $6^+$ states with $\it{v}$$=4$ are situated far in energy. The calculated B$(E2;2^+\rightarrow0^+)$ value also remains almost unchanged after the modification in the interaction. Since a reduction in the $f_{7/2}$ matrix elements can be related to a reduction in neutron pairing, we may infer that the n-n pairing appears to be only marginally reduced for these neutron-rich nuclei.

Our calculated B$(E2;6^+\rightarrow4_1^+)$ value in $^{138}$Sn also increases from $12.3$ $e^2fm^4$ to $16.8$ $e^2fm^4$ on modifying the interaction, and comes close to the experimental value of $16(4)$ $e^2fm^4$. The $4_2^+$ state, which was $7$ keV below the yrast $6^+$ state, now lies $94$ keV above it, making the mixing of two $4^+$ states less likely. This suggests that the $6^+$ isomer is following almost pure seniority $\it{v}$$=2$ scheme in $^{138}$Sn. The other B$(E2;4_1^+\rightarrow2^+)$ and B$(E2;2^+\rightarrow0^+)$ values also change from $0.5$ to $17.9$ $e^2fm^4$ and $188.7$ to $163.3$ $e^2fm^4$, respectively. No experimental data for these B$(E2)$ values are available for comparison.

The structure of yrast states in $^{134}$Sn turns out to be much simpler. Only two valence neutrons are available which occupy the $f_{7/2}$ orbital. We note that a pure $\it{v}$$=2$ seniority scheme is applicable for the yrast $2^+$, $4_1^+$ and $6^+$ states in $^{134}$Sn. The level energies and the B$(E2)$ values can be explained reasonably well by the ``RCDB'' interaction, which remain nearly unchanged even after using the modified ``RCDBMO'' interaction. The calculated B$(E2;6^+\rightarrow4_1^+)$ and B$(E2;2^+\rightarrow0^+)$ values agree with the experimental data as shown in Fig.~\ref{fig:level}. This implies that no core excitations are required to explain the structure of $^{134}$Sn. This also supports the earlier claims that N$=82$ is a robust shell closure~\cite{dworschak}.

Fig.~\ref{fig:occu} shows the average occupancy of the $f_{7/2}$ and $p_{3/2}$ orbitals for the yrast $4_1^+$ and $6^+$ states along with the $4_2^+$ state in $^{134-138}$Sn isotopes. The yrast $4_1^+$ and $6^+$ states dominantly consist of the $f_{7/2}$ orbital in all the isotopes. However, the $4_2^+$ state dominantly consists of the $f_{7/2}$ orbital in the $^{136}$Sn isotope, while it has a significant mixing of the $p_{3/2}$ orbital with the $f_{7/2}$ orbital in the $^{134,138}$Sn isotopes.

Also, the $4_2^+$ state lies below to the $6^+$ isomeric state in the $^{136}$Sn isotope with both the ``RCDB'' as well as the ``RCDBMO'' interactions. The occurrence of the $4_2^+$ state, therefore, can strongly affect the B$(E2;6^+\rightarrow4_1^+)$ value in the case of $^{136}$Sn. We note that the two $4^+$ states coming from $f_{7/2}$ orbital can not have the same seniority. The yrast $4_1^+$ state is most likely to be a $\it{v}$=2 seniority state while the second $4_2^+$ state is most likely to be a $\it{v}$=4 seniority state and the two can mix with each other. The seniority mixing seems to be responsible for a nonzero B$(E2;6^+\rightarrow4_1^+)$ value in the $^{136}$Sn isotope, as already discussed. The B$(E2;6^+\rightarrow4_1^+)$ value increases from $14.5$ $e^2fm^4$ to $22.0$ $e^2fm^4$, while the B$(E2;6^+\rightarrow4_2^+)$ value decreases from $53.1$ $e^2fm^4$ to $46.4$ $e^2fm^4$, on modifying the interaction (See Fig.~\ref{fig:level}). The relative energy gap between the two $4^+$ states also slightly decreases after modification of the interaction. It seems that the seniority mixing in both the $4^+$ states increases by reducing the $f_{7/2}$ matrix elements, as they come closer in energy with almost similar average occupancies.

 On the other hand, the relative gap of the two $4^+$ states increases in the $^{138}$Sn isotope so that the $4_2^+$ state lies above the yrast $6^+$ isomeric state on using the ``RCDBMO'' interaction. There is a slight decrement in the average occupancy of the $f_{7/2}$ orbital, and increment in the average occupancy of the $p_{3/2}$ orbital for the $4_2^+$ state with the modified interaction, while the opposite happens in the yrast $4_1^+$ and $6^+$ states (See Fig.~\ref{fig:occu}). The B$(E2;6^+\rightarrow4_1^+)$ value, therefore, increases and comes close to the experimental value, because of the less mixing of the two $4^+$ states. The $4_2^+$ state lies far from the yrast $4_1^+$ state in $^{134}$Sn, making the mixing of two $4^+$ states very unlikely, as already discussed above with the applicable $\it{v}$=2 pure-seniority scheme for the yrast $4_1^+$ and $6^+$ states.

\section{Conclusions}
We conclude that the ``RCDB'' interaction appears to work reasonably well for the highly neutron-rich $^{134-138}$Sn-isotopes lying beyond the doubly closed $^{132}$Sn with a small reduction of $25$ keV in the ${\nu}f_{7/2}^2$ matrix elements. This is in contrast to the large reduction of $150$ keV required in the calculations by Simpson $\it{et}$ $\it{al.}$~\cite{simpson}. Further, we do not find any need to include the contribution from the core excitations even in $^{134}$Sn. We also conclude that a pure $\it{v}$$=2$ seniority scheme is more applicable in $^{134}$Sn and $^{138}$Sn while a seniority mixing is seen in $^{136}$Sn; this leads to the observed mismatch in the B$(E2;6^+\rightarrow4_1^+)$ value in $^{136}$Sn. A modification in the interaction leads to a larger B$(E2;6^+\rightarrow4_1^+)$ value because of increased seniority mixing in the $4_1^+$ and $4_2^+$ states of $^{136}$Sn.

We also note that further refinement in our calculated results may be possible if we decide to reduce the diagonal and off-diagonal matrix elements differently. We, however, confirm that a small modification in the interaction beyond the N$=82$ magic number is indeed required. We may also conclude that the ``RCDB'' interaction seems better suited to explain the data of the neutron rich systems in the N$=82-126$ region. It would be very useful to measure the remaining B$(E2)$ values for the transitions in $^{134-138}$Sn-isotopes in order to gain further understanding of the effective interactions in the neutron-rich region.

\begin{acknowledgments}
Financial support from Department of Science and Technology, Department of Atomic Energy, and Ministry of Human Resource Development (Govt. of India) is gratefully acknowledged.
\end{acknowledgments}
\newpage 
\bibliography{apssamp}

\end{document}